\documentclass[prap,twocolumn,superscriptaddress,noshowpacs,nobalancelastpage]{revtex4}
\usepackage{color}
\usepackage{float}
\usepackage[pdftex]{graphicx}
\usepackage{graphicx}
\usepackage{dcolumn}
\usepackage{bm}
\usepackage{siunitx}
\usepackage{comment}
\usepackage{fixltx2e}
\usepackage{soul}

\usepackage{hyperref}
\hypersetup{
    bookmarks=true,         
    unicode=false,          
    pdftoolbar=true,        
    pdfmenubar=true,        
    pdffitwindow=false,     
    pdftitle={Charge detection in an array of CMOS quantum dots},    
    pdfauthor={Dr. Matias Urdampilleta},     
    pdfsubject={},   
    pdfcreator={Dr. Matias Urdampilleta},   
    pdfproducer={},  
    pdfkeywords={,} {} {}, 
    pdfnewwindow=true,      
    colorlinks=false,       
    linkcolor=red,          
    citecolor=green,        
    filecolor=magenta,      
    urlcolor=cyan           
}

\usepackage{filecontents}

\DeclareSIUnit{\belmilliwatt}{Bm}
\DeclareSIUnit{\dBm}{\deci\belmilliwatt}
\begin{document}
\bibliographystyle{apsrev4-1}

\preprint{APS}
\title{Charge detection in an array of CMOS quantum dots}

\author{Emmanuel Chanrion}
\email{emmanuel.chanrion@neel.cnrs.fr}
\affiliation{Univ. Grenoble Alpes, CNRS, Grenoble INP, Institut N\'eel, 38402 Grenoble, France}

\author{David J. Niegemann	}
\affiliation{Univ. Grenoble Alpes, CNRS, Grenoble INP, Institut N\'eel, 38402 Grenoble, France}

\author{Benoit Bertrand}
\affiliation{CEA, LETI, Minatec Campus, F-38054 Grenoble, France}

\author{Cameron Spence}
\affiliation{Univ. Grenoble Alpes, CNRS, Grenoble INP, Institut N\'eel, 38402 Grenoble, France}

\author{Baptiste Jadot}
\affiliation{Univ. Grenoble Alpes, CNRS, Grenoble INP, Institut N\'eel, 38402 Grenoble, France}

\author{Jing Li}
\affiliation{Univ. Grenoble Alpes, CEA, IRIG, 38000 Grenoble, France}

\author{Pierre-Andr\'e Mortemousque}
\affiliation{CEA, LETI, Minatec Campus, F-38054 Grenoble, France}

\author{Louis Hutin}
\affiliation{CEA, LETI, Minatec Campus, F-38054 Grenoble, France}

\author{Romain Maurand}
\affiliation{Univ. Grenoble Alpes, CEA, IRIG, 38000 Grenoble, France}

\author{Xavier Jehl}
\affiliation{Univ. Grenoble Alpes, CEA, IRIG, 38000 Grenoble, France}

\author{Marc Sanquer}
\affiliation{Univ. Grenoble Alpes, CEA, IRIG, 38000 Grenoble, France}

\author{Silvano De Franceschi}
\affiliation{Univ. Grenoble Alpes, CEA, IRIG, 38000 Grenoble, France}

\author{Christopher B{\"a}uerle}
\affiliation{Univ. Grenoble Alpes, CNRS, Grenoble INP, Institut N\'eel, 38402 Grenoble, France}

\author{Franck Balestro}
\affiliation{Univ. Grenoble Alpes, CNRS, Grenoble INP, Institut N\'eel, 38402 Grenoble, France}

\author{Yann-Michel Niquet}
\affiliation{Univ. Grenoble Alpes, CEA, IRIG, 38000 Grenoble, France}

\author{Maud Vinet}
\affiliation{CEA, LETI, Minatec Campus, F-38054 Grenoble, France}

\author{Tristan Meunier}
\affiliation{Univ. Grenoble Alpes, CNRS, Grenoble INP, Institut N\'eel, 38402 Grenoble, France}

\author{Matias Urdampilleta	}
\email{matias.urdampilleta@neel.cnrs.fr}
\affiliation{Univ. Grenoble Alpes, CNRS, Grenoble INP, Institut N\'eel, 38402 Grenoble, France}

\date{\today}
\begin{abstract}
The recent development of arrays of quantum dots in semiconductor nanostructures highlights the progress of quantum devices toward large scale. However, how to realize such arrays on a scalable platform such as silicon is still an open question. One of the main challenge resides in the detection of charges within the array. It is a prerequisite functionality to initialize a desired charge state and readout spins through spin-to-charge conversion mechanisms. In this paper, we use two methods based on either a single-lead charge detector, or a reprogrammable single electron transistor. Thanks to these methods, we study the charge dynamics and sensitivity by performing single shot detection of the charge. Finally, we can probe the charge stability at any node of a linear array and assess the Coulomb disorder in the structure. We find an electrochemical potential fluctuation induced by charge noise comparable to that reported in other silicon quantum dots. 
\end{abstract}

\maketitle

\section{Introduction}
The ability to create qubits with long lifetime out of single electron spins in lithographically defined quantum dots has turned silicon into a promising platform for quantum computing. 
The long coherence time of spins embedded in isotopically enriched silicon combined with an improved control of coherent manipulation allows for high-fidelity single and two-qubit gates \cite{Zajac2018, Watson2018, yoneda2018quantum, Huang2019}. 
Moreover, recent developments have shown that spin readout can be performed with high fidelity and short timescale \cite{Zheng2019, PhysRevX.9.041003, Urdampilleta2019, DzurakReflecto}, making error correction codes possible in scalable architectures \cite{li2018crossbar, Veldhorst2017, hutin2018quantum}.
Following this progress, the next milestone is to increase the number of qubits under control while preserving the above mentioned performance. 

In parallel, 2D arrays of quantum dots have been recently explored \cite{Mukhopadhyay2018}, in which basic quantum functionalities\cite{Mortemousque2018} as well as condensed matter simulations\cite{Dehollain2019} have been demonstrated. However, these demonstrations have been achieved in GaAs heterostructures where the hyperfine interaction limits the coherence time to a few tens of ~\si{\nano\second}. In order to create functional quantum dot arrays on a more scalable platform, such as silicon quantum dots \cite{PhysRevApplied.6.054013, Lawrie2020, Betz2016}, the same level of control and adressability needs to be achieved.

In this paper, we report a first important step toward this goal. We demonstrate the remote charge sensing of quantum dots in an architecture that contains 8 MOS quantum dots (QDs), see Fig. \ref{fig:device}. Each QD is operated with a single gate electrode: the experiment is carried out on a 4 split-gate n-type silicon device, fabricated on a silicon-on-insulator 300-mm wafer using an industry-standard fabrication line. 
We propose two detection-schemes based on a charge sensor embedded in the device that allows us to probe the charge stability and dynamics in the different QD configurations as well as the Coulomb disorder in the structure. Moreover, these detection methods allow us to assess the static and dynamical Coulomb disorder which is an important characterization step to understand and improve the quality of qubit devices.

The first charge detection method consists in using a single lead quantum dot (SLQD) at one end of the array, probed by radiofrequency reflectometry \cite{PhysRevApplied.6.044016}. 
This method shows a very high charge sensitivity to the first and second neighbors as well as high single-shot charge readout fidelity (99.9$\%$ at 1kHz). 
We can envision to use this SLQD as a readout site in quantum protocols exploiting spin shuttling at one end of the array for qubit readout \cite{Fujita2017, Mills2019} or in 3D structures \cite{Batude, C3NR33738C}. 
The second method relies on a reconfigurable single electron transistor (SET) \cite{Lawrie2020}. For this purpose, one can exploit any of the QDs of the upper or lower linear array.
It allows to sense any quantum dot in the other parallel array and shows, as well, high single shot charge fidelity. 

The paper is organized as follows. 
We first describe how the devices are designed and operated. In a second time, we discuss how the SLQD- and SET-based detector work and how we operate them to probe the stability diagrams of the neighbouring double quantum dots in the few electrons regime
Then, we exploit these two methods to investigate single-shot detection of charge tunneling in the array and the dynamical and static Coulomb disorder inside the device.
Finally, we conclude by discussing how these two sensing methods could be used in a protocol to form a logical qubit in a 1D array of quantum dots \cite{PhysRevX.8.021058}.

\section{Device design and operation}

\begin{figure}%
\includegraphics[width=\columnwidth]{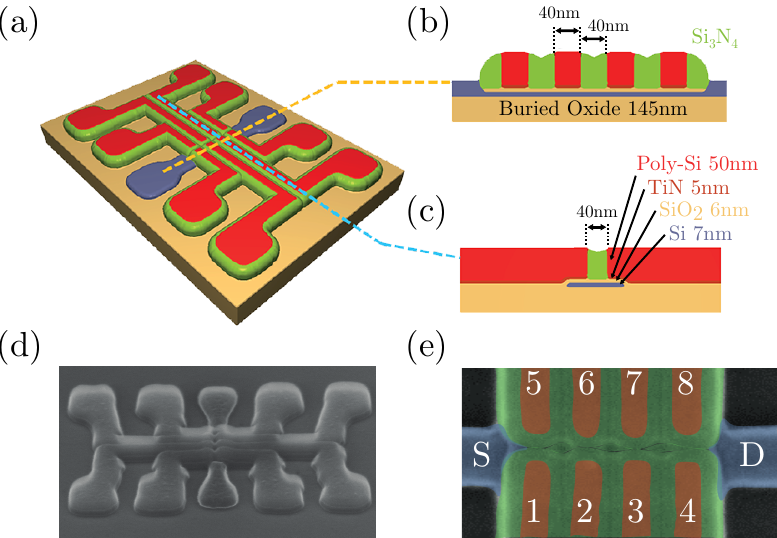}
\caption{ 
(a) Picture of the 2x4 QD array. The silicon nanowire (blue) is covered with top gates (red) which are separated by spacers (green). The non-covered regions of the nanowire are highly doped to form electron reservoirs. (b) Cross section along the nanowire. (c) Cross section along one top gate. (d) SEM micrograph of a device similar to the ones used in the present study. (e) False color SEM micrograph of the array using the same color code as (a). The quantum dots (QD1 to QD8) are localized below the top gates (G1 to G8) in the corners of the nanowire. 
}
\label{fig:device}%
\end{figure}
The devices, such as the one depicted in Fig. \ref{fig:device}, were fabricated on $300$~\si{\milli\metre} SOI substrates (Buried oxide thickness $T_{BOX}$= $145$~\si{\nano\metre}), see Fig. \ref{fig:device}(b). The silicon channel (width W=$70$ to $110$~\si{\nano\metre}) was defined by mesa patterning. The gate stack (see Fig. \ref{fig:device}(c)) was made of $6$~\si{\nano\metre} thermally grown SiO$_{2}$, $5$~\si{\nano\metre} ALD-deposited TiN, $50$~\si{\nano\metre} of Poly-Si, and topped by a bilayer hard mask (HM) with $30$~\si{\nano\metre} SiN and $25$~\si{\nano\metre} SiO$_{2}$.
A hybrid DUV/EBeam Gate patterning scheme was implemented, in which multiple lithography-etch cycles were performed sequentially to transfer parts of the final pattern into the HM, prior to a final transfer etch.
The resulting structure consists of 4 pairs of split gates along a silicon nanowire channel, each overlapping opposite edges of the mesa. The gate pitch along the nanowire direction is $80$~\si{\nano\metre} (gate with length $40$~\si{\nano\metre}, spaced by $40$~\si{\nano\metre}), and the split width is $40$~\si{\nano\metre}.
The doped areas are defined in a self-aligned way, outside of regions covered by the gates and an offset spacer. Thus, a particularly wide ($35$~\si{\nano\metre}) Si$_3$N$_4$ offset spacer was deposited, completely covering the inter-gate spacings. The Si areas still exposed were regrown by means of epitaxy, before undergoing ion implantation of n-type dopants activated by a N$_2$ spike anneal. This regions form the electron reservoirs, labeled S and D (source and drain) in the figures by analogy with classical MOS devices.

At dilution fridge temperature (40mK), quantum dots can be formed at the Si/SiO$_{2}$ interface by applying a positive voltage on the top gates, as previously reported. This allows to form up to 2x4 quantum dots, one dot underneath each gate electrode, see Fig. \ref{fig:device}(e). 

To sense single charges in the nanostructure, we use two different methods. The first one consists in using a SLQD as a charge detector. It is positioned next to a reservoir at the left end of the gate array and is probed by radio-frequency gate-reflectometry. For this purpose, the gate controlling its electrochemical potential is connected to a tank circuit formed by an inductance and the parasitic capacitance to ground. Therefore, the phase of the reflected signal is sensitive to a change in quantum capacitance. The phase shift exhibits a Coulomb-like peak when a level of SLQD is in resonance with the Fermi sea of the lead. To sense the charge occupancy of neighbouring dots, the SLQD is biased on the side of such a peak. 
In this position, no current flows through the device due to large negative voltages applied to the gates next to the drain (gates 3, 4, 7 and 8), efficiently closing the barrier to that reservoir.
Any change in the electrostatic environment induces a shift of the Coulomb peak and thereby alters the reflected RF-signal. As will be developed in the next section, this method allows us to sense the first and second neighbours of the SLQD. 

The second charge sensing method consists in operating the upper side of the 2x4 array as a SET. For this purpose, three out of the four upper gates are set to 1.2V which corresponds to a regime where these quantum dots act as electron reservoirs. The last gate controls the SET which is operated in the many electrons regime. Similarly to the previous method, the embedded SET is biased on a Coulomb peak to reach the maximum sensitivity. It is worth noting that in order to minimize tunneling effects between the SET and the probed quantum dot, we have implemented this method in a wide wire device ($110$~\si{\nano\metre}).    

\section{Charge detection in the array}

\subsection{Charge sensing using RF-single-lead quantum dot}

\begin{figure}%
\includegraphics{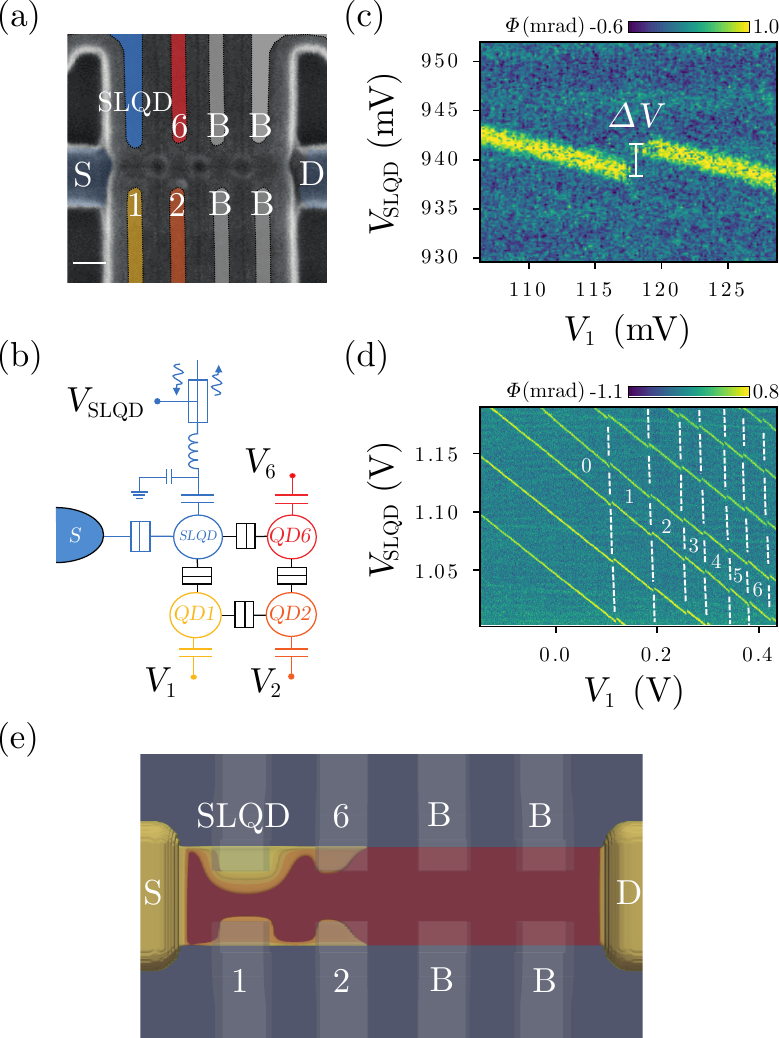}%
\caption{(a) False color SEM micrograph of the QDs configuration in the SLQD detection mode. B stands for barrier gates where the voltage is set to -1V. (b) Equivalent electrical circuit of the device probed by dispersive readout. One quantum dot is used as an electrometer (SLQD) and is tunnel and capacitively coupled to the three neighbouring QDs. Its gate is connected to an inductance to form an LC resonant circuit which is probed by RF-reflectometry. (c) The phase change of the resonant circuit is plotted as a function of $V_{\mathrm{SLQD}}$ and $V_1$. The signal lines correspond to a charge degeneracy of the electrometer dot which experiences a shift in voltage for one electron added to QD1. The voltage shift corresponds to 2.4 linewidths of the detector. (d) Charge stability diagram of QD1. (e) Simulation of the electron density inside the channel under the same polarization conditions as the experiment and using the Thomas-Fermi approximation. A large density of electron is present at the SLQD location which overlaps with the QD6 potential. A large dot accumulates under the SLQD gate, and tends to spill over QD6 to which it is strongly coupled.}
\label{fig:dispersive}%
\end{figure}

To demonstrate charge sensing by dispersive readout, we first focus on the 2x2 subarray depicted in Fig. \ref{fig:dispersive}(a). It consists of one QD used as an electrometer and three QDs that are probed. The other gates are set to -1V to decouple the array from the drain reservoir.
In this configuration, we use QD5 as an electrometer and re-label it QDE to distinguish it from the probed QDs (QD1, QD2 and QD6). 
We connect the gate controlling the SLQD to an inductance $L=820~\si{\nano\henry}$ to achieve dispersive sensing of the electrometer, as depicted in Fig. \ref{fig:dispersive}(b). 
Together with the parasitic capacitance to ground $C_p=0.39$ ~\si{\pico\farad}, we obtain an LC circuit with a resonant frequency f=286MHz and a quality factor Q=70. 
The signal once demodulated gives amplitude and phase shift of the reflected radiofrequency wave. 
The maximum phase shift is measured on the top of a Coulomb peak and is directly related to the quantum capacitance, $C_Q$, $\Delta\Phi\simeq 2Q\times C_Q /C_p$ \cite{petersson2010charge}. 
We obtain $\Delta\Phi=1$mrad and $C_Q=12$aF in average for the peak shown in Fig. \ref{fig:dispersive}(c).

Figure \ref{fig:dispersive}(c) shows such a Coulomb peak as a function of the gate voltage controlling QD1. The slope of the detector against $V_1$ is due to capacitive coupling between QDE and the gate G1. The addition of one charge in QD1 induces an abrupt voltage shift $\Delta V$ on the position of the electrometer Coulomb peak. Fig. \ref{fig:dispersive}(d) shows the extended charge stability diagram where the filling of QD1 up to 6 electrons is observed. 

\begin{figure}%
\includegraphics{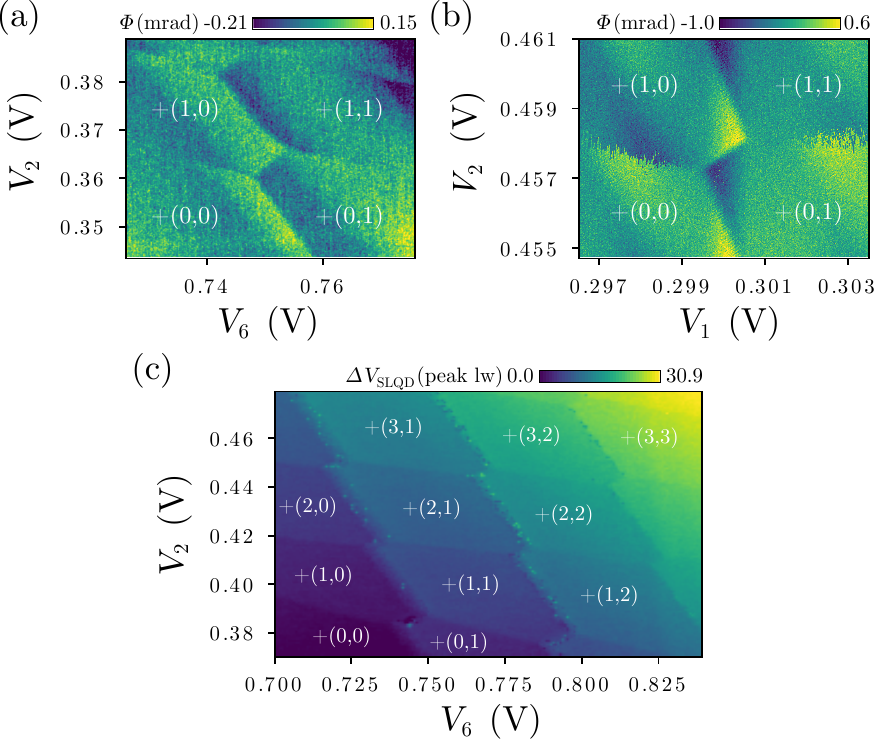}%
\caption{(a) and (b) Stability diagrams of the “QD2-QD6” and “QD1-QD2” DQDs. The dispersive signal from QDE is plotted as a function of ($V_2$, $V_6$) and ($V_1$, $V_2$). (c) Stability diagram of the QD2-QD6 DQD on extended charge configurations. We plot the shift in detector position (normalized by the peak linewidth) induced by a change in charge occupancy as a function of $V_2$ and $V_6$. 
}
\label{fig:compensate}%
\end{figure}

The shift of the SLQD chemical potential induced by charging events is larger than the Coulomb peak linewidth. Since the detector sensitivity vanishes outside this peak, a feedback loop assisted detection is impractical in this case. 
However, it is possible to tune the gate voltage in order to sense the double quantum dots formed by (QD1, QD2) and (QD2, QD6). 
Fig. \ref{fig:compensate}(a) and (b) shows the stability diagram of the 2 DQDs in the few electron regime. 
It is important to note that the shift induced by a charge in QD6 (4.7 linewidth) is stronger than in QD2 or QD1(1.1 and 2.4 linewidth respectively).
This can be explained by the localization of the quantum dot in the corner of the nanowire \cite{doi:10.1021/nl500299h}. Therefore, the capacitive and tunnel coupling along the nanowire axis is stronger than along the transverse axis. To confirm this hypothesis, we have computed the carrier density in a self-consistent Thomas-Fermi approximation. The source and drain are assumed doped with $N_d = 10^{20}$ phosphorous per cm$^3$, whose ionization probabilities are calculated with an incomplete ionization model valid at low temperature \cite{Altermatt2006}. The density of electrons, $n(\textbf{r}) = N_c F_{1/2}[(E_c-eV(\textbf{r})-\mu)/(kT)]$, depends on the local potential $V$(\textbf{r}), which includes the mean-field contribution from the ionized impurities and electrons themselves ($F_{1/2}$ being the Fermi integral, $N_c$ and $E_c$ the effective density and states and conduction band edge energy in bulk silicon, and $\mu$ the chemical potential). The calculations were run at $T$ = 20K.  Although the Thomas-Fermi approximation does not account for quantum effects such as confinement and tunneling, it is expected to give a fair account of the position of the dots and transport channels in the system. Figure \ref{fig:dispersive}(e) shows that a large dot does form under the SLQD gate, as expected. It tends to spill under the neighboring QD6 gate while it remains fairly decoupled from the facing QD1 and QD2 dots, which is in good qualitative agreement with the experiments.
The sensitivity of the detector is limited from one to a few charge transitions before the signal vanishes. A working sensing position can only be held for voltage offsets $\Delta V_1, \Delta V_2 \sim 20\si{\milli\volt}$, due to small detector linewidth and strong capacitive coupling.
However, the coarse tuning of a double-dot like (QD1, QD2) implies to explore a voltage space $(V_1, V_2)$ that can be as large as $1\si{\volt}\times1\si{\volt}$.
To facilitate this tuning despite the absence of feedback loop, we calibrate the SLQD detector over the whole voltage space. 
For this purpose, we measure the shift in SLQD chemical potential at the boundaries of this voltage space and we interpolate the optimum window for the detector for each point assuming a constant capacitive coupling model.
This method makes the sensor sensitive over the whole probed region as illustrated by Fig \ref{fig:compensate} (c) which shows a stability diagram of the (QD6, QD2) DQD over many charge configurations. 
It is also possible to probe extended QD combinations as illustrated by supplemental materials S1 and S2 where triple-dot charge stabilities are explored.

\subsection{Charge sensing using embedded SET}
\begin{figure}%
\includegraphics[width=\columnwidth]{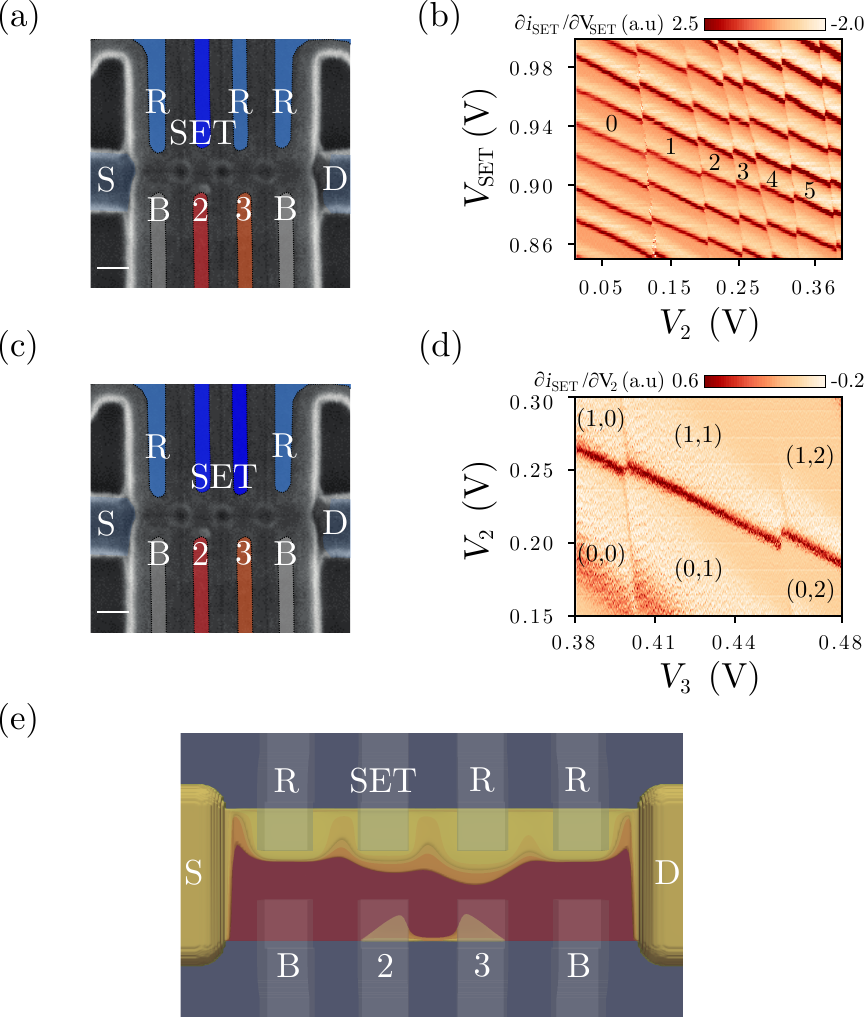}%
\caption{
(a) False color SEM micrograph of the QDs configuration in the SET detection mode. B stands for barrier (V=-1V) and R for reservoir (V=1.2V). 
(b) Charge stability diagram of QD2. The derivative of the current through the device is plotted as a function of $V_{\mathrm{SET}}$ and $V_2$. In this case, the QD in front of QD2 is operated as a SET while the other upper gates are set to high positive voltages ($>1.2\si{\volt}$) to extend the reservoirs close to the SET.
(c) False color SEM micrograph of the QDs configuration in the double SET detection mode. The top linear array of quantum dots is operated as two SETs in series. The Coulomb blockade is probed by measuring the current flowing through the structure.
(d) Stability diagram of a (QD3, QD2) DQD in the few electrons regime. The detector is operated in the large bias regime ($3\si{\milli\volt}$) to extend the region of sensitivity without the need for capacitive compensation. (e) Iso-density surfaces inside the channel computed in the Thomas-Fermi approximation at the following gate bias: $V=\SI{-1}{\volt}$ on the barrier gates,  $V=\SI{1.2}{\volt}$ on the reservoir gates, $V=\SI{0.8}{\volt}$ on the SET and $V=\SI{0.3}{\volt}$ on QD2 and QD3. As the bias is larger on the reservoir gates than on the SET, and owing to the cross-capacitance between the upper and lower gates, the density in the SET reservoirs tends to spill toward QD3. This shall further decrease the sensitivity of the SET to changes in the occupation of QD3.
}
\label{fig:current}%
\end{figure}
To probe the charge and spin configuration of QDs located deeper in the array, the above solution requires to shuttle electrons to one of the neighbouring dots, as the SLQD is not sensitive to QDs 3, 4, 7 and 8. However, such protocol requires a fine control of tunnel barriers at fast time scales. Another approach we use here is to operate the upper part of the array as a reconfigurable SET. For this purpose, all the gates but one are set to accumulation mode ($>1.2\si{\volt}$) to extend the reservoirs as close as possible to the SET. We check that we can form a SET with the last gate left by measuring a Coulomb map which exhibits a regular pattern of Coulomb diamonds (see supplemental materials B). Figure \ref{fig:current} (c) presents the stability diagram of QDE and QD2 probed with this technique. Similarly to the SLQD method, one can clearly see a shift in voltage on the detector for every electron accumulated in QD2.

To sense a multi-QD system such as the (QD2-QD3) DQD, forming a single SET localized under one gate is not very efficient. The presence of the SET reservoir screens the capacitive coupling to other QDs across the wire which limits the sensitivity of the SET to a single QD. This is supported qualitatively with the simulation presented in Fig. \ref{fig:current}(e). It shows that the accumulation of electrons under one of the reservoir gate expands inside the channel toward one of the facing QD. Therefore, we use another strategy which consists in forming SETs in series along the nanowire axis, Figure \ref{fig:current} (c). They are operated in the tunnel broadened and large bias ($>3\si{\milli\volt}$) regime in order to lift Coulomb blockade and have a finite current flowing through the SETs at any voltage. Fig. \ref{fig:current}(d) shows the (QD2, QD3) DQD stability diagram in the few electron regime probed by two serial SETs. However, we operate the double SET in a regime where Coulomb oscillations are less sharp than in the case of a single SET, which decreases the sensitivity.

\section{Probing single charge tunneling and Coulomb disorder }

\subsection{Time-resolved charge detection}
The readout of spin qubits in semiconductor quantum dots is achieved by mapping the spin information on the charge degree of freedom, through energy selective tunneling \cite{Elzerman2004}, tunnel rate selective tunneling \cite{PhysRevLett.94.196802} or Pauli spin blockade between two QDs \cite{ono2002current, PhysRevLett.103.160503}. Hence, the ability to perform high fidelity single shot readout of a single charge tunneling is a key requirement for quantum computation using spin qubits. Here, we demonstrate the single shot readout of charge using the two detection methods described earlier.

\begin{figure}%
\includegraphics[width=\columnwidth]{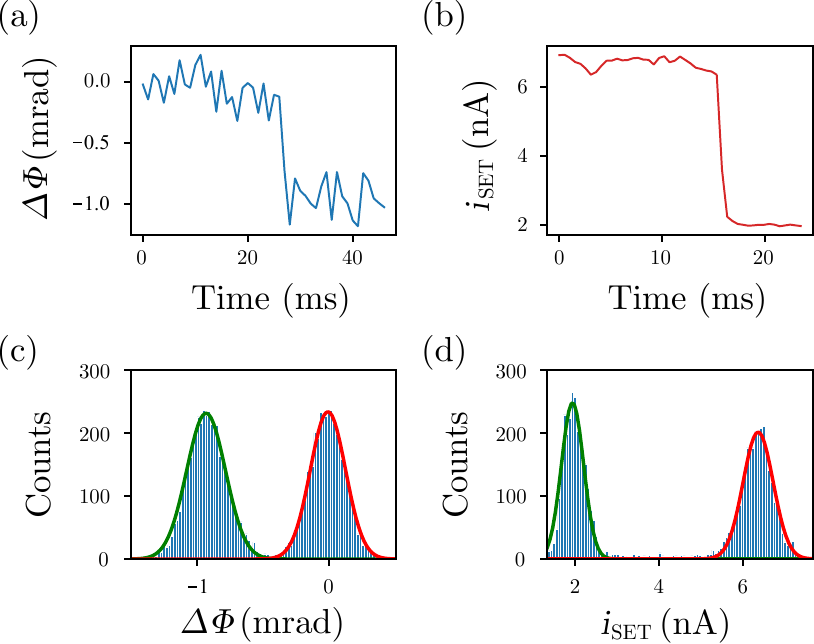}%
\caption{(a) and (b) Time-resolved measurement of a single charge tunnelling out of the QD2 probed by dispersive (resp. current-based) measurement for $1\si{\milli\second}$ integration time. (c) and (d) Histogram of 8000 single shot measurements with an integration of $1\si{\milli\second}$. The SNRs are 7 and 15 respectively.
}
\label{fig:fidelity}%
\end{figure}

To isolate a charge tunneling event in the measurement bandwidth ($1\si{\kilo\hertz}$), we focus on QD2 which is only slightly coupled to the reservoir when $V_1$ is set to $0\si{\volt}$. The slow tunneling of the charge, already evidenced by the non-continuous delineation between charge configuration (0,0) and (1,0) for instance (see Fig. \ref{fig:compensate}(b)), is measured as a function of time using the two different detection methods, see Fig. \ref{fig:fidelity} (a) and (c). On these two graphs, we can observe the charge tunneling in the QD2 while the detector experiences a shift in phase or in current.

To analyse the signal-to-noise ratio (SNR), we first load an electron on QD2 then pulse QD2 chemical potential above the Fermi sea. We perform that way 8000 single shot measurements at a time where there is equal probability to have the QD2 loaded or empty. We integrate the signal for $1\si{\milli \second}$ and plot the resulting phase shift as a histogram, see Fig. \ref{fig:fidelity}(b) and (d). The data are fitted with two Gaussian curves that allow to extract the theoretical error in discriminating between the two charge states. The minimal error rate on a charge assignment is $10^{-3}$ for the SLQD based readout and $10^{-7}$ for the embedded single SET based readout. 
Moreover, the SNR gives a time domain charge sensitivity of $5.0\times 10^{-3}$ e/$\sqrt{\si{\hertz}}$ and $2.1\times 10^{-3}$ e/$\sqrt{\si{\hertz}}$ respectively. These figures are comparable to what has been reported in literature \cite{PhysRevLett.110.046805,PhysRevApplied.6.054013} but could be enhanced.
In the case of dispersive readout, the maximum phase shift is given by $\Delta\Phi_{max}\simeq Q\times (e\alpha)^2/(2tC_p)$ with $\alpha$ the lever arm and $t$ the tunnel coupling between the lead and SLQD. Hence, the signal could be improved by controlling the tunnel coupling using an extra gate or by increasing the capacitive coupling between the SLQD and its top gate. 
The noise on our measurement is estimated to be around $0.1\si{\nano\volt/\sqrt{\hertz}}$ which is equivalent to a noise temperature of $4.5\si{\kelvin}$ and corresponds to the noise of the cryogenic amplifier used here. Noise could be reduced by using a superconducting amplifier such as a Josephson parametric amplifier \cite{MortonJPA} which would decrease the noise temperature by more than one order of magnitude. 
In the case of the embedded SET-based measurement the amplitude of the signal can be increased by using modern transimpedance amplifiers which are now operational at dilution temperatures \cite{8036394}.

\subsection{Low-frequency charge noise and static Coulomb disorder}
The two methods we have developed allow to measure the charge dynamics of single electron tunneling in quantum dots but can also be used to probe the charge noise within the structure. 
Charge noise may be the ultimate source of decoherence in isotopically purified silicon, due to finite spin-orbit coupling or to the presence of inhomogeneous magnetic field \cite{yoneda2018quantum}. 
It is therefore crucial to understand its origin to improve the single- and two-qubit gates fidelities. 

\begin{figure}
\includegraphics[width=\columnwidth]{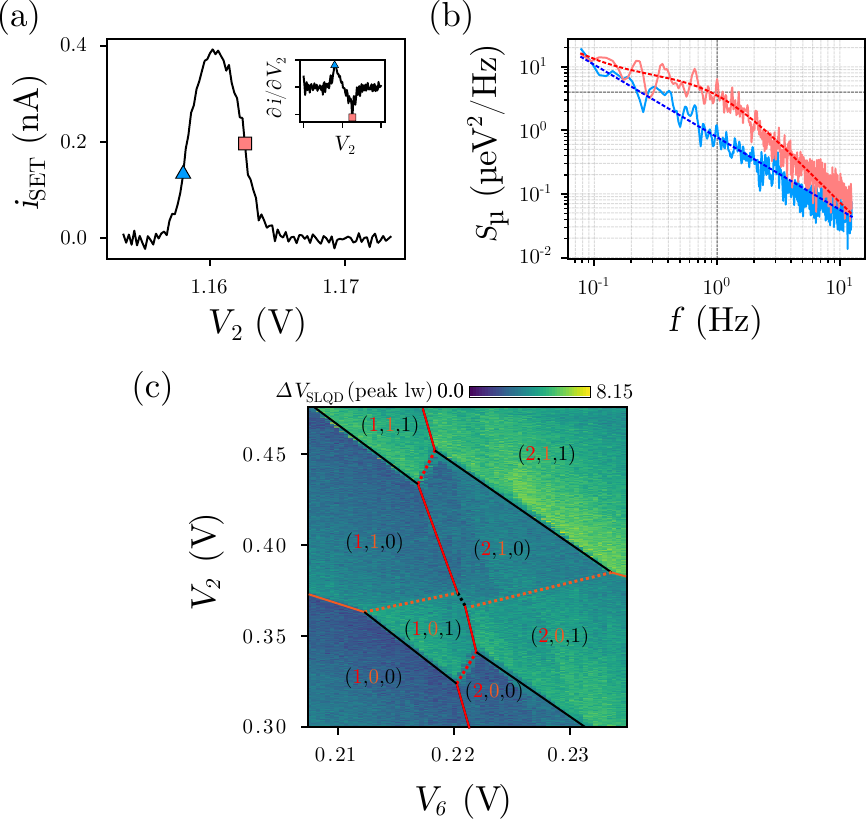}%
\caption{(a) Coulomb peak of the SET located under $V_2$. Inset: derivative of current against the gate voltage. (b) Power spectral density of the noise in chemical potential on the two sides of the Coulomb peak at 350mK. The blue curve is fit using a 1/f$^{\beta}$ model ($\beta$=1.14) and the red curve with a function of the form $\frac{A}{f^{\beta}}+\frac{B}{f^2/f_c^2+1}$ ($\beta$=1.4 and $f_c$=1Hz). (c) Stability diagram of the triple dot system formed by QD2, QD6 and a defect in the channel.
}
\label{fig:noise}%
\end{figure}

In this context, we operate the top array as a single SET to assess the charge noise inside the structure. We then measure the noise spectrum on the chemical potential induced by charge fluctuations, $S_{\mu}$, at two different points of the SET Coulomb peak, depicted in Fig. \ref{fig:noise}(a). At these points the slope is maximum giving access to the highest charge sensitivity. Figure \ref{fig:noise}(b) presents the two corresponding charge noise spectra, $S_{\mu}$. The first spectrum shows a 1/$f^{\beta}$, with $\beta=$1.14, which is closed to the 1/f  behavior expected for a uniform distribution of fluctuating  two-level systems (TLSs) in the environment \cite{Paladino2019}. The second spectrum shows a stronger deviation from a 1/f noise. To explain this, we use a more refined model of charge noise which accounts for a non-uniform distribution of TLS activation energies compared to the temperature \cite{Dutta1979, Connors2019}. $S_{\mu}$ is fit using a function of the form $S_{\mu}=\frac{A}{f^{\beta}}+\frac{B}{f^2/f_c^2+1}$. The Lorentzian shape visible on Fig. \ref{fig:noise}(b) suggests the presence of a single fluctuator whose characteristic frequency, $f_c$, is centered around 1Hz. However, the amplitude of the potential fluctuation at 1Hz, is around 3.5$\mu$eV/$\sqrt{Hz}$, which indicates that the TLS is only slightly coupled to the quantum dot. For comparison, the lowest potential fluctuations values reported in literature for silicon are in the 2-5 $\mu$eV/$\sqrt{Hz}$ range at 350mK \cite{Connors2019, Kim2019, Mi2018, Freeman2017}.

While mobile charges can affect spin qubit coherence, fixed charges can be detrimental for the production of QDs with low dot-to-dot variability.
Indeed, the presence of static disorder modifies the electrostatic environment of the QD. We can probe the formation of unintentional quantum dot by charge sensing. 
Figure \ref{fig:noise}(c) shows an example of a stability diagram measured between QD2 and QD6. 
The presence of an accidental quantum dot is clearly visible as the expected honey-comb pattern is not visible. 
Instead, we obtain different charge configurations when the QD2 and QD6 exchange an electron with the defect. 
This kind of signatures are visible on few of the stability diagrams (see Supplemental Materials) and can be attributed to unintentionnal doping of the channel during the formation of the source and drain. 
To avoid the presence of these fixed charges in the channel, other fabrication routes are currently under investigation such as the epitaxial formation of reservoirs.

\section{Conclusion}
In conclusion, we have shown the ability to build a device with a large number of quantum dots capacitively and tunnel coupled. The charge occupancy of any of the quantum dots in the structure can be probed either using a reconfigurable SET or a SLQD. Moreover, we have demonstrated that these two methods offer the ability to perform real time measurement of a single charge tunneling event which could be used to load single electrons in the structure or readout spin states. Finally, we exploit these methods to assess the static and dynamical Coulomb disorder in the structure.

One possible way to operate a quasi-1D array of QDs consists in using the upper row as qubits and the lower row as ancillas for readout \cite{7838409}. In this configuration, we propose to combine the two detection methods presented here to initialize and readout the array. To initialize the two rows with one electron per dot, we first use the lower array as a SET to monitor the loading of 2 electrons per dot in the upper array. Then, the SET is turned off and one electron is transferred to each lower array QD. To readout each qubit in the upper array, we can first convert the spin information into charge by using Pauli spin blockade on the ancilla dot. Once the information stored in the charge occupancy (either 1 or 2 charges in each lower dot) the charges can be shuttled to the end of the array where they will be sensed using the SLQD detector. Such operation requires a good control of the array charge occupancy and a fine control of tunnel barriers \cite{Mortemousque2018, Ansaloni2020}.

\section{Acknowledgment}
We acknowledge technical support from P. Perrier, H. Rodenas, E. Eyraud, D. Lepoittevin, I. Pheng, T. Crozes, L. Del Rey, D. Dufeu, J. Jarreau, J. Minet and C. Guttin. 
D.J.N. and C.S. acknowledges the GreQuE doctoral programs (grant agreement No 754303). The device fabrication is funded through the Mosquito project (G. A. 688539) and QuCube (G. A. 810504)
This work is supported by the Agence Nationale de la Recherche through the MAQSi, CMOSQSPIN and CODAQ projects (ANR-16-ACHN-0029).

\clearpage
\renewcommand{\figurename}{Supplementary Figure}
\setcounter{figure}{0}
\section*{Supplementary material}
\subsection{Triple quantum dots in the few-electron regime using SLQD and SET based sensing.}\label{SUPP-1}
Both SLQD and reconfigurable SET based methods allow us to probe the multi-dot system in the few electron regime. Figure S1 (a) and (b) show charge stability diagrams of a DQD (two different devices). We can clearly see on the two diagrams the presence of an unintentionnal QD in the channel. It is similarly coupled to both gates implying that it is located in the center of the channel or at the top interface between the gates.
\begin{figure}[H]%
\includegraphics{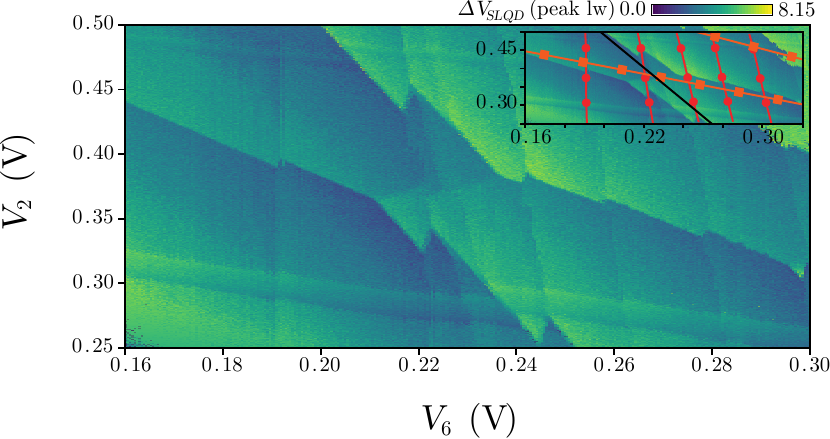}%
\caption{
Stability diagram of a multidot system in the nanowire probed using a SLQD. We attribute each transition to either QD1 (red), QD2 (orange), or to an additional unintentionnal QD (black). A zoom in the triple dot regime at the center is available in the main text.
}
\end{figure}
\begin{figure}[H]%
\begin{center}
\includegraphics[scale=1]{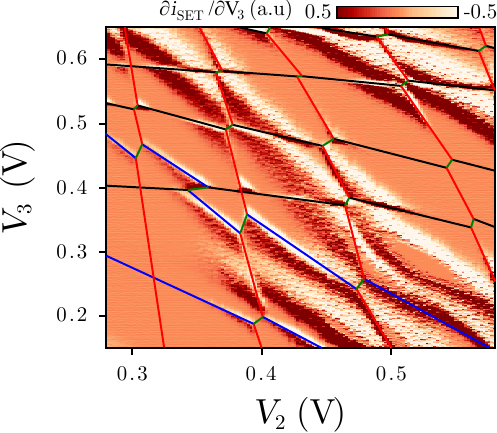}%
\end{center}
\caption{Stability diagram of a multidot system in the nanowire probed by reconfigurable SET.
}
\end{figure}
\subsection{Coulomb map of a single QD SET.}\label{SUPP-3}
We demonstrate that a single dot, from the upper part of the array, can be used as an SET while the bottom part remains in the blockade regime. 
The figure S4 shows an SET formed by QD7 where we can clearly see a regular pattern of Coulomb diamonds.
\begin{figure}[H]%
\begin{center}
\includegraphics{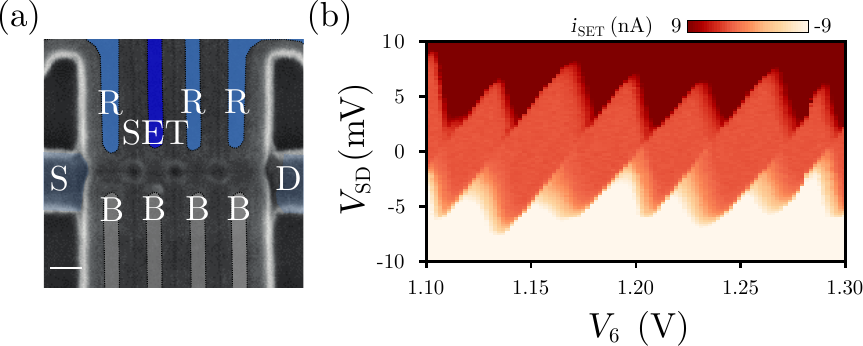}%
\end{center}
\caption{
(a) False-color SEM picture of the device. The upper QD array is operated as a reconfigurable SET.
(b) Coulomb map of the SET formed by QD7. 
}
\end{figure}

\subsection{Stability diagram of quantum dots in the lower array probed by SET-sensing.}\label{SUPP-2}
Here, we present the charge stability diagram for each quantum dot in the bottom linear array, except for QD2 which is presented in the main text.
\begin{figure}[H]%
\begin{center}
\includegraphics{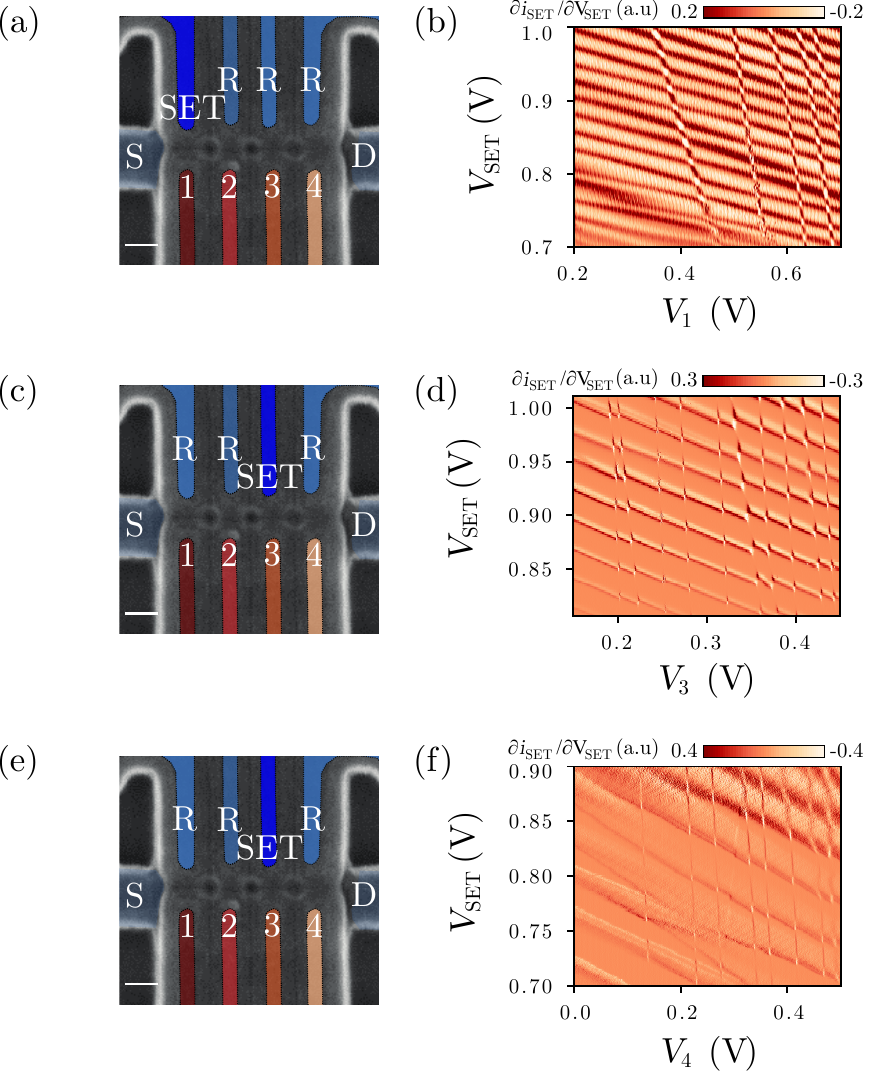}%
\end{center}
\caption{(a), (c), (e) False-color SEM picture of the device. The upper QD arrays is operated as a reconfigurable SET. (b), (d) and (f) are stability diagrams of the lower array QDs 1,2 and 4 measured using the above dot as a SET. The irregular pattern in addition energy is a signature of a strong Coulomb disorder.
}
\end{figure}

\end{document}